# Nanocrystalline grain boundary engineering: Increasing Σ3 boundary fraction in pure Ni with thermomechanical treatments

David B. Bober[1,2], Mukul Kumar[2], Timothy J. Rupert[1]*

[1] Department of Mechanical and Aerospace Engineering, University of California, Irvine, CA 92697, USA

[2] Lawrence Livermore National Laboratory, Livermore, CA 94550, United States

*E-mail: trupert@uci.edu

**Abstract**

Grain boundary networks should play a dominant role in determining the mechanical properties of nanocrystalline metals. However, these networks are difficult to characterize and their response to deformation is incompletely understood. In this work, we study the grain boundary network of nanocrystalline Ni and explore whether it can be modified by plastic deformation. Mechanical cycling at room temperature did not lead to structural evolution, but elevated temperature cycling did alter the grain boundary network. In addition to mechanically-driven grain growth, mechanical cycling at 100 °C led to a 48% increase in Σ3 boundaries, determined with transmission Kikuchi diffraction. The extent of boundary modification was a function of the number of applied loading cycles and the testing temperature, with more cycles at higher temperatures leading to more special grain boundaries. The results presented here suggest a path to grain boundary engineering in nanocrystalline materials.

## Introduction

Nanocrystalline metals are promising next-generation structural materials with high strength, fatigue life and wear resistance [1-3]. Their enhanced properties can be attributed to the high density of grain boundaries, which is usually quantified indirectly by grain size. Indeed, grain size has been the fundamental metric used for the creation of nanocrystalline structure-property scaling laws to this point [4]. However, recent studies have highlighted the importance of also considering boundary type and topological arrangement [5-7]. Nanotwinned Cu, which contains grains subdivided into nanoscale twin domains, is perhaps the most notable example. This material shares the enormous strength of nanocrystalline copper but remains ductile because twins replace random boundaries as the dominant network component, providing soft and hard directions for dislocation movement [5, 8]. In essence, the properties are improved by substituting a favorable boundary type (twin) for an unfavorable one (random).

Atomistic modeling studies suggest that changes to the grain boundary network can alter mechanical behavior. Rupert and Schuh [9] observed that subtle boundary relaxation, either through annealing or mechanical cycling, could increase the strength of a simulated nanocrystalline metal. Since catastrophic strain localization can occur if a high strain path percolates across a nanocrystalline specimen [10], ductility should also be intimately connected to features of the boundary network. Hasnaoui et al. [6] used molecular dynamics to show that shear strain can concentrate in the random boundaries, which resist sliding less strongly than low angle boundaries. Experimental studies support these observations, with boundary relaxation found to increase strength but also promote shear localization [11, 12]. In addition to mechanical properties, it has become evident that grain boundary network characteristics are closely tied to the thermal stability of nanostructured materials. Lagrange et al. [7] showed that a few highly mobile boundary segments can cause coarsening in an otherwise stable network. Clearly, the exact character and arrangement of grain boundaries is critical to the performance of nanostructured metals, providing motivation to study and control nanocrystalline grain boundary networks.

Tuning grain boundary networks is accomplished in coarse grained alloys through grain boundary engineering (GBE) treatments [13]. Most commonly, this consists of repeated thermomechanical processing to maximize the number and connectivity of boundaries which are considered "special" [13]. Several analytical tools exist to classify grain boundaries and characterize a boundary network. Most GBE investigators apply the coincident site lattice (CSL) model, which is based on the maximum theoretical periodicity of shared lattice sites [14]. The CSL values ($\Sigma$) can be approximately correlated with boundary energy [14]. Boundaries with $\Sigma$ value less than 29 are considered special because of their low energy. Low angle ($\Sigma 1$) and twin ($\Sigma 3$) boundaries are often singled out for specific consideration because of their unique properties [13]. Triple junctions can be categorized according to the number of special ($\Sigma 1$-29) boundaries which they join. Network connectivity, which controls many intergranular phenomena, has been evaluated with both percolation theories and triple junction distributions [15, 16]. More recently, network topology has been quantified with the cluster mass approach, which is based on the length of interconnected boundary segments sharing a common type [17].

Typical GBE treatments are of two types, often called strain annealing and strain recrystallization [18]. In strain annealing, the metal is deformed 6-8% and then heated below the recrystallization temperature for several hours, with the whole process repeated several times [19, 20]. Unfortunately, the long annealing times cause significant grain growth and increase processing costs [18]. Alternatively, in strain recrystallization, the sample is deformed 5-30% and then heated to a high temperature for a short time, with these steps being iterated as needed [21]. In strain recrystallization, the level of strain energy is insufficient to cause complete recrystallization upon heating, instead causing boundary decomposition [18]. When this occurs, a boundary is split into new segments by the nucleation of a grain [18]. A higher energy boundary will tend to decompose into multiple lower energy segments, reducing the total system energy [18]. Inhomogeneous strain energy density causes the nucleated grain to expand into its neighbors, elongating the new low energy boundary segments [18]. When repeated, this produces a fine-grained microstructure in which special boundaries are well incorporated into the network [18].

Traditional GBE mechanisms should not operate in nanocrystalline systems, primarily because these materials cannot store the dislocation networks required to drive boundary decomposition. The lack of dislocation storage in nanocrystalline metals has been shown by in situ X-ray diffraction experiments, transmission electron microscopy (TEM) investigations, and atomistic simulations [22-24]. Fortunately, nanocrystalline metals deform through collective processes, which may supply a replacement mechanism for grain boundary network evolution. At grain sizes below ~100 nm, dislocations are emitted from grain boundaries and absorbed at interfaces on the other side of the grain [1]. While the traditional view of a dislocation is that it brings a small increment of plasticity, a single dislocation moving through a 15 nm grain in Al can cause a shear strain of ~2%, enough to change the grain shape [25, 26]. Such deformation should also require accommodation from surrounding grains to maintain compatibility at the interfaces. Grain rotation and sliding are the predominant carriers of plastic strain for grain sizes below ~20 nm [27, 28]. The underlying physical processes behind such mechanisms are atomic shuffling events at the boundary, which some authors have likened to the shear transformation zones in metallic glasses [24, 29, 30]. A priori, grain rotation must alter the local grain boundary character since the misorientation of the interface is changing. The possibility of longer range boundary modification is suggested by observations of textured nanocrystalline clusters formed by stress driven rotation [31]. The high stresses accessible in nanocrystalline materials can also lead to grain boundary migration, causing grain growth, softening, and increased ductility [32-34]. The common feature of these phenomena is that they may directly modify the boundary network, providing the potential to controllably modify nanocrystalline grain boundary networks.

Until recently, the study of nanocrystalline grain boundary network reorganization was impossible due to the limitations of characterization tools. Electron backscatter diffraction (EBSD), the standard tool for studying coarse-grained networks, lacks sufficient resolution to characterize nanocrystalline materials. Two new techniques have been developed which relieve this difficulty, namely TEM based automated crystal orientation mapping (TEM-ACOM) and scanning electron microscope (SEM) based transmission Kikuchi diffraction (TKD) [35, 36]. TEM-ACOM uses precession enhanced convergent beam electron diffraction and pattern matching to enable orientation mapping with ~2 nm resolution [36]. A limitation of this technique for studying nanocrystalline materials is that diffraction patterns are generated for every grain the beam passes through. If there are overlapping grains, the resulting signals may be impossible to deconvolve [36]. This places a very strict limit on sample thickness and may pose challenges for specimen preparation. TKD uses Kikuchi patterns cast on an EBSD detector by electrons forescattered through a thin specimen [35]. The low interaction volume enables resolution down to ~3 nm, although this is dependent on atomic number [35]. The Kikuchi pattern is generated from only the lowest surface of the sample and thickness limits are significantly relaxed [37]. This technique also has the major advantage of being implementable on a standard EBSD equipped field emission SEM.

With new characterization techniques and an improved understanding of nanocrystalline deformation physics, the tools are now available for a study of nanocrystalline interfacial networks and their evolution under stress. In this study, we explore methods to directly modify the grain boundary network of nanocrystalline Ni, using combinations of applied stress and elevated temperature. Room temperature mechanical cycling was found to be ineffective, leaving the grain structure and boundary network unchanged. However, cycling at elevated temperature did induce evolution of the grain boundary network. Grain boundary network evolution was most obviously observed as an increase in the Σ3 boundary fraction. The effects of stress-free annealing and of creep were separately investigated to provide controls with which to compare the other treatments, showing that high stress and plastic deformation are needed for microstructural evolution. The most likely mechanism is believed to be collective deformation, although our results focus on statistical boundary metrics rather than micromechanisms.

## Materials and methods

Nickel was selected for these experiments because its high stacking fault energy makes it challenging to grain boundary engineer by traditional means [38]. Nanocrystalline Ni was deposited onto Si wafers from a 99.999% pure Ni target using radio frequency magnetron sputtering for a final thickness of 260±9 nm (Ulvac JSP 8000). All of the material used for this work was deposited in a single batch to ensure a uniform as-deposited condition. The supporting Si was micromachined to produce a rigid frame around a free-standing Ni membrane, with dimensions of 2.5 by 10 mm. The fabrication process generally followed Vlassak et al., with deep reactive ion and $XeF_2$ etching replacing KOH and reactive ion etching, respectively [39].

The thin films were mechanically loaded with bulge testing, where a gas pressure deforms a diaphragm-like sample [40]. The test is convenient for thin film materials because the specimen and supporting window are co-fabricated, which eliminates sample handling and gripping issues [39]. It is also less cumbersome than microtensile or microcompression tests, especially when testing at elevated temperatures, while still delivering full stress-strain measurements. A custom bulge test apparatus, which is capable of performing controlled thermomechanical cycling at temperatures up to 500 ˚C, was constructed. The device was similar to that of Kalkmann et al. [41] , except that deflection was measured with a standard laser triangulation sensor (Micro-Epsilon optoNCDT 1700).

Bulge testing produces a biaxial stress, much like in the thin wall of a cylindrical pressure vessel [39]. For long rectangular specimens, the hoop stress and strain are constant across the sample, allowing for the uniform onset of plasticity within the film [42]. The hoop strain ($\varepsilon_1$ ) is given by:

$$\varepsilon_1 = \varepsilon_0 + \frac{a^2+h^2}{2ah} arcsin\left(\frac{2ah}{a^2+h^2}\right) - 1 \quad (1)$$

where $p$ is the applied pressure, $a$ is ½ the membrane width, $\varepsilon_0$ is the residual strain and $h$ is the maximum bulge height [42]. The hoop stress ($\sigma_1$) is given by:

$$\sigma_1 = \frac{p(a^2+h^2)}{2ah} \quad (2)$$

Many authors who employ bulge testing combine this hoop stress with the smaller longitudinal stress component to calculate a von Mises equivalent stress. However, at elevated temperatures, some of our films become slack before testing. While they become taut in the hoop direction as soon as a small pressure is applied, making Eqns. (1) and (2) valid as long as the initial height is properly considered [43], the films take longer to become taut in the longitudinal direction. To remain consistent, we only report hoop stresses and strains in this study. While this can affect measurements of mechanical properties such as strength, calculating these properties is not our primary goal. Rather, repeatable mechanical cycling is our aim.

The bulge tests were controlled by setting the pressure as a function of time. According to Eqn. (2), the stress cannot be controlled without knowledge the bulge height. Rather than implement a complex feedback system, the peak cycling pressure was selected from our monotonic tests on identical samples. If the peak deflection remains nearly constant, then this method will produce cycles with roughly constant stress amplitude, approximating stress-controlled cycling. In the case of accumulated plastic strain, the peak stress will drop as the peak bulge height increases. The applied pressure was varied slowly, at a cycling rate of 4 mHz, producing an average strain rate on the order of $10^{-6}$ $s^{-1}$.

Specimens were prepared for TKD and plan view TEM by thinning at cryogenic temperatures in a low angle Ar ion mill (Fischione 1010) at 2-3 kV and 5 mA. Immediately prior to TKD analysis, the samples were cleaned in a 10 W oxygen plasma for 5 minutes (South Bay Technologies PC2000). Cross-sectional TEM specimens were prepared using the focused ion beam (FIB) in situ lift-out technique in a Quanta 3D field emission gun (FEG) dual beam microscope. A voltage of 5 kV was used during the final thinning step to minimize the thickness of the damaged layer created by the FIB. Bright field TEM images were collected with an FEI/Philips CM-20 instrument operated at 200 kV. TKD was performed with an FEI Quanta3D equipped with an Oxford Instruments Nordlys F+ EBSD detector, operated at 30 kV and 11 nA with a 1 mm aperture and a 3.5 mm working distance. These parameters were selected using the information published by Keller and Geiss [35], and Trimby and coworkers [44, 45]. A custom holder was used to align the sample at a 20° tilt to the beam axis, as in Keller and Geiss's geometry [35]. A step size of 2-6 nm was selected based on the grain size. Maps were kept small to minimize drift, usually from 5000-24000 $nm^2$. A representative example of the as-collected TKD maps is shown in Figure 1(a).

Orientation data was processed and standard noise reduction techniques were applied [46, 47] using the Channel5 software package (Oxford Instruments). A 2° critical misorientation angle was used to reconstruct grain boundaries and the Brandon criterion was then applied to categorized them into CSL types [48], as shown in Figure 1(b). Each boundary segment was classified as either random high angle or special ($\Sigma \leq 29$). Of the special boundaries, extra attention was given to the low angle ($\Sigma$1), twin ($\Sigma$3), and twin variant ($\Sigma$9, $\Sigma$27) types. These types were selected because they can be correlated with boundary energy and are commonly reported in traditional GBE research. Grain boundary statistics are reported according to length fraction because this measure is less sensitive to short, erroneously indexed boundaries [49]. Triple junction types were identified using code written in MATLAB (MathWorks). The grain size was calculated from the reconstructed grain areas, as recommended by ASTM E2627 [50]. The ASTM E2627 provision to discard grains with fewer than 100 indexed points was impractically restrictive, and was lowered to a threshold of 4 points [50].

## Results and Discussion

Bright field TEM images of the as-deposited microstructure are shown in Figure 2. The plan view image in Figure 2(a) shows uniformly nanocrystalline and equiaxed grains, having a mean size of 23 nm. There are no abnormally large grains and the grain size distribution appears narrow. Material with a small grain size was desired to maximize the potential for grain boundary network reorganization through collective deformation physics. Vo et al. [51] showed that the amount of plastic strain which could be attributed to grain rotation is inversely related to grain size, theoretically increasing the overall network modification with decreasing grain size. Below grain sizes of ~10 nm, grain boundary sliding and rotation can become so dominant as to cause an inverse Hall-Petch effect [52]. A lower limit on the practical grain size was imposed by the resolution of TKD, which could only reliably detect grains larger than 5 nm. As such, a 23 nm mean grain size provided a good compromise between small grain size and TKD data quality. TKD measurement gave an average grain size of 22 nm, providing excellent agreement with the TEM results. All figures in this paper which quote grain size are referring to measurements taken from TKD.

A cross-sectional TEM micrograph of an as-deposited film is presented in Figure 2(b), showing only modest grain elongation in the film normal direction (ND). Many boundaries are perpendicular to the deposition direction, making this structure distinct from the columnar morphology sometimes found in sputtered films. An equiaxed grain structure is desirable for this study because it should mimic the response of a truly bulk nanocrystalline material. Free surface effects on stress-driven grain boundary migration are generally limited to a region within a distance of about one grain diameter from the surface [53]. In our samples, the presence of many grains through the film thickness limits the importance of any free surface effects.

The as-deposited film texture is represented by the pole figures shown in Figure 3, which demonstrates a slight texture in the film's growth direction. Of the total material, 34±6% was oriented within 15° of the ideal <111> normal direction fiber (<111>//ND). This fiber texture is common to many sputtered films [54, 55]. In general, fiber textures will increase the fraction of CSL boundaries which share the same misorientation axis [56]. For a <111> fiber texture, an increased fraction of Σ 1, 3, 7, 13b, 19b and 21a boundaries would be expected [56]. We observed no measurable change in film texture after any combination of mechanical cycling or annealing.

Samples were first deformed at room temperature to investigate our hypothesis that grain boundary mediated plastic deformation can alter the boundary network. In addition to monotonic loading to failure, a cyclic loading pattern was also used because it is expected to cause greater microstructural changes. Stress induced grain coarsening, one obvious form of boundary evolution, has been observed in several studies of nanocrystalline Ni deformed at room temperature. The magnitude of coarsening averaged ~600% higher in those studies that applied cyclic loads [2, 57-64], although it is impossible to control for differences such as sample purity and loading type. Molecular dynamics work has also linked cyclic stress to grain boundary evolution, showing a reduction in local structural disorder with increasing number of cycles [9]. Figure 4(a) shows the hoop stress-strain curves for films loaded both monotonically and cyclically. The nanocrystalline films demonstrate the high strength and low ductility characteristic of most nanocrystalline metals. A high peak cyclic stress of 1.2 GPa was selected to maximize the potential for stress driven grain boundary migration and grain rotation. The modulus was measured to be 150 GPa, and did not vary significantly between loading and unloading or with extended cycling. The brittle nature of the films did not allow a yield strength to be determined. Despite the lack of a clear yield point, plastic strain accumulated over the course of several cycles, eventually reaching a maximum of 0.35%. Isolation of the 1st, 5th, 25th and 125th cycles (presented in Figure 4(b)) shows that the initial hysteresis quickly disappears, indicating that any microstructural change is concentrated in the first few cycles.

The gradual drop in peak stress with increased cycling is caused by the open-loop pressure controlled test procedure, as previously discussed.

Since grain coarsening is an obvious sign of structural evolution, we begin by looking at the grain size distributions. Figure 5(a) shows that the as-deposited, monotonically loaded, and cyclically loaded materials have identical mean grain size and distribution. No grain coarsening occurred, unlike some other reports on fatigue loaded nanocrystalline Ni [2]. Possible subtle changes to the grain boundary network were assessed by analyzing TKD orientation maps. The grain boundary character distribution (GBCD), which tracks CSL fractions, is shown in Figure 5(b). All error bars in this work represent a 95% binomial proportion confidence interval and were calculated using the Normal Approximation Method. Neither monotonic nor cyclic loading caused any change in the GBCD. Together, the unchanged grain size and GBCD indicate that no significant boundary network change was driven by mechanical cycling at room temperature. This may be due to the limited plasticity; all of the literature motivating our hypothesis involved significantly more plasticity than was achieved in this case.

Analytical models indicate that the amount of grain rotation should increase with increasing temperature, following an Arrhenius trend. Cahn and Taylor described rotation as the combined result of coupled grain boundary motion and sliding [65]. In the case of only coupled grain boundary motion, the overall extent of rotation will be directly tied to boundary mobility, until very high temperatures when the coupling breaks down [65, 66]. The contribution of grain boundary sliding to grain rotation has been isolated and modeled by Moldovan [67], who followed Raj and Ashby's [68] work. They showed that the sliding rate depends on the lattice and grain boundary diffusivities, which, like boundary mobility, follow an Arrhenius relation. Harris et al. [69] reached an identical conclusion using an analysis built on Ashby and Verrall's work [70]. Atomistic simulations by Panzarino et al. [71] agree with these analytical models, showing that stress driven grain rotation is enhanced at high temperature. Experimental work has also shown that elevated temperature can promote stress driven nanocrystalline grain growth [72, 73].

Therefore, to encourage plasticity and concomitant structural evolution, the testing temperature was increased. For this study, the ideal temperature would allow for significant boundary mobility without causing thermal grain growth. Films were annealed at 100, 150, and 250 ˚C to gain a better understanding of the effect of temperature on thermal grain growth. At 150 and 250 ˚C, abnormal grain growth consumed most of the nanocrystalline material. At 100 ˚C, a modest level of limited abnormal grain growth occurred, but most of the material remained nanocrystalline. This seemed to provide the best balance between boundary mobility and grain size stability.

The stress strain results for mechanical testing at 100 ˚C are shown in Figure 6(a), where the monotonic behavior and cyclic response are both shown. For clarity, the 1st, 5th, 10th and 50th cycles have also been plotted separately in Figure 6(b). The peak stress amplitude was reduced to 900 MPa to avoid rupturing the sample. Elevated temperature reduced the modulus to 110 GPa and increased the total plastic strain to nearly 1%. It is apparent that the plastic strain accumulated exceeds that observed during room temperature cycling. The mid-loop hysteresis and incremental increase in plastic deformation decreased with increasing cycle number. Taken together, the cycling behavior indicates more potential for microstructural change than was seen at room temperature, despite the lower peak stress.

The top row of Figure 7(a) presents TEM images from samples that were mechanically cycled at 100 ˚C, showing that cycling was accompanied by grain growth. The extent of microstructural change qualitatively correlates with the number of loading cycles. To isolate the effect of cyclic plasticity, a series of stress-free annealing experiments was also performed. In Figure 7, each image of cycled material is above an image of material which was annealed stress-free for a time matching the duration of the cycling experiment. The annealed specimens show much less structural evolution, indicating that cyclic plasticity is of prime importance. The bottom row of Figure 7 also shows that annealing at 100 °C can cause a few large grains to form. This abnormal grain growth is expected for pure nanocrystalline metals and is a mechanism for reducing excess boundary energy [74, 75]. To preserve this study's focus on nanocrystalline phenomena, the abnormally grown grains were excluded from the TKD analysis. Ideally, this would have been done by collecting very large maps and then post-process sorting by grain size. Unfortunately, drift required that maps be so small that they could not encompass the larger grains, which would have made post-process sorting ambiguous. It was instead decided to exclude the large grains prior to orientation mapping by selecting nanocrystalline regions based on the forescatter image.

Grain size distributions are presented in Figure 8 for both the stress-free and mechanical cycling treatments at 100 °C. For stress-free annealing, the grain size distribution remains unchanged through 100 min. A small increase in grain size is observed after annealing for 250 min. On the other hand, mechanical cycling caused significant changes to the grain size distribution. Fifty loading cycles increased the average grain size to 51 nm, up from a starting value of only 22 nm, and the entire grain size distribution shifts toward the largest values. After 50 cycles, the microstructural evolution was so extensive that it became slightly more difficult to identify and exclude the abnormally grown grains using only the forescatter image. This reduces our certainty that the 50 cycle data is entirely free from the influence of larger grains. In any case, the trend for increasing grain size with mechanical cycling remains obvious. Similar grain growth caused by repeated mechanical stress has previously been observed in several studies, as mentioned previously [2, 57-64]. The observed coarsening could be caused by stress driven boundary migration or rotation induced coalescence [76, 77].

The coarsening trend indicates grain boundary rearrangement, the nature of which was investigated with TKD orientation mapping. Four types of grain boundaries are quantified in Figure 9, as a function of applied mechanical cycle. Data from the stress-free annealing control experiments is also included to provide a baseline. The largest change in the grain boundary character distribution was an increase in Σ3 fraction, shown in Figure 9(b). The small change after 1 and 5 cycles was followed by increases of 30% after 10 cycles and 48% after 50 cycles. In each case, the change is quantified relative to the starting material. The trend of increasing Σ3 fraction with increased cycling is further evidence that grain boundary rearrangement is driven by the repeated deformation. The Σ3 length fraction remains constant during the stress-free annealing.

There are currently very few reports which statistically quantify the grain boundary network of nanocrystalline materials because the required microscopy techniques are so new. In one such work, Brons and Thompson [78] have reported the grain boundary character distribution for a sputtered Ni film created under slightly different processing conditions. For an initial grain size of 37 nm and strong <101>//ND texture, these authors found a Σ3 length fraction of only 5.2% [78]. The Σ3 length fraction increased to a maximum of 9.2% after annealing at 450 °C, which also caused substantial coarsening. In comparison to these results, the 34.4% Σ3 fraction which we observe here appears to be quite high. The difference in film texture may contribute to the difference between the results of Brons and Thompson and our own.

Kobler et al. [79] found that deformation could either increase or decrease the number of twins per grain in nanocrystalline Pd. They found that twin density fell in samples with a large initial concentration of twins, while it rose in material which initially contained few twins per grain [79]. This suggested to them that deformation was driving the sample toward an equilibrium state which balanced concurrent twinning and detwinning [79]. In support, Kobler et al. cited prior reports showing that nanocrystalline metals can twin and detwin under deformation, even for high stacking fault materials [80, 81]. Luo et al. [82] found that the number of twinned grains in nanocrystalline Au increased after fatigue loading, although their proposed twin assisted grain growth mechanism appears to predict twins with misorientations outside the accepted range. Figure 10 shows the cumulative distribution of twins based on their deviation from the ideal 60° <111> misorientation, up to the Brandon criterion of 8.66°. In such a figure, perfect twins will skew the distribution towards the left and the curve becomes sharper, while less perfect twins will cause it to skew right and more gradually rise to the total twin fraction present. Mechanical cycling causes the distribution to skew progressively leftwards, toward low deviation angles. This reveals that cycling preferentially increases the number of near-perfect twins. Such a trend suggests that there is an increase in the fraction of coherent twins, which are more likely to have near-perfect misorientations [83]. This rearrangement could be facilitated by the rotation of existing twins toward perfect twinning, as reported in molecular dynamics experiments by Panzarino et al. [71]. It could also be explained by an increase in the length or number of annealing twins. It is desirable to better quantify the types of $\Sigma 3$ boundaries because of the dramatically different properties they may exhibit [83].

Unfortunately, the two-dimensional TKD data does not provide boundary plane inclination. The stereological method developed by Saylor et al. [84] is not applicable because of the low number of boundaries. In the future, emerging three-dimensional techniques with nanometer resolution could be used to provide added detail [85].

Our as-deposited material had an initial Σ1 fraction of 5.8%, much lower than the 25.6% reported by Brons and Thompson [78]. Changes to the Σ1 fraction are shown in Figure 9(a), revealing that mechanical cycling at 100 °C caused the Σ1 fraction to decrease to 2.3%. One possible explanation is the grain-rotation-coalescence model proposed by Haslam et al. [77], in which rotation reduces boundary misorientation until neighboring grains merge. Panzarino et al. also observed that neighboring grains could rotate and coalesce into new grains with bent lattices [71]. Fatigue stress in ultrafine grained Cu has similarly reduced the fraction of low angle grain boundaries (LAGB) [86].

The fractions of Σ9 and Σ27 boundaries is slightly reduced by mechanical cycling (although always within the error bars of our annealed data), which is in contrast to the behavior encountered in coarse-grained grain boundary engineering. An increase in the Σ3 fraction typically leads to more Σ3-Σ3 interactions, which in turn produce twin variants by the CSL product rule [87]. The slight drop in Σ1 and Σ9, 27 fractions yielded a total special boundary fraction (Σ1-29) that increased less than the Σ3 fraction. The CSL fractions discussed here have been included in Table 1 for easy reference.

Triple junction distributions were used to quantify the frequency of interactions between special and random boundaries. In a typical GBE process, type 3 triple junctions would be expected to increase and type 0 junctions would decrease, which is indicative of increasing special-special interactions [15, 87]. The effect of annealing and mechanical cycling on the fraction of each junction type is plotted in Figure 11. The small number of junctions sampled led to wide confidence limits and requires a cautious interpretation. The type 3 junctions underwent the expected increase, showing a special boundary fraction that is incorporating into the grain boundary network [15]. The slight drop in type 2 boundaries can be explained by their conversion to type 3 boundaries under the triple junction product rule. The unchanged type 1 fraction is also predicted by the theoretical triple junction distribution [15]. The constant type 0 fraction differs from the theoretically predicted drop [15]. Overall, the trend is suggestive of an increasing Σ3 fraction that is somewhat integrated into the boundary network. The small map size used in this study precludes corroborating this with a more rigorous cluster mass analysis of network connectivity [17].

The structural evolution we observe in the TEM images, cumulative grain size distribution functions, and GBCDs all correlate with the number of stress cycles. However, before attributing the cause of these observations to cyclic plasticity, the possible role of creep needs to be explored [2, 88]. To this point, creep and cyclic effects could have been conflated because mechanical cycling exposes the specimen to high stress for a time which is proportional to the number of cycles. To isolate these phenomena, another specimen was cycled at 8 mHz (twice the usual frequency) for 50 cycles. This halved the total time the specimen was exposed to high temperature and stress (125 min versus the original 250 min), while keeping the number of cycles unchanged. Any creep effects should therefore be more pronounced in the sample cycled at low frequency, i.e. for a longer time. Figure 12 shows that the GBCD is insensitive to the duration of cycling, demonstrating that creep is of negligible impact and that cyclic plasticity is the driving mechanism. An in situ TEM fatigue study of nanocrystalline films by Kumar et al. [89] showed that grain rotation during cycling can be caused by reversible dislocation motion. If this is the case, it is likely that a ratcheting mechanism can reduce the overall boundary energy [9]. Panzarino et al. [71] used molecular dynamics to show that grain sliding and rotation can also result in increased levels of microstructural evolution as the number of loading cycles is increased. In both of these studies, cyclic loading modified the microstructure in ways not observed under monotonic loading. The relative importance of these several mechanisms and their dependence on thermomechanical conditions are currently open questions. Atomistic simulations will hopefully provide a definitive answer in the near future.

It remains to be shown what effects the observed microstructural changes may have on physical, chemical or electrical properties. The slight changes in triple junction fractions, along with the relatively low total special boundary fraction, suggest that the connectivity of the random boundary network will not be disrupted [15]. This implies that intergranular degradation will not be reduced, even though the Σ3 fraction increased significantly. Still, the increased special boundary fraction would likely affect other properties. Specifically, grain boundary sliding is strongly affected by boundary type; low energy boundaries being most shear resistant and acting to concentrate shear along random ones. Hasnaoui et al. [6] showed that this could produce localized shear flows between special boundary clusters. In addition, dislocation nucleation and dislocation-GB interactions are strongly affected by GB structure [89]. In the case of nanotwinned metals, special dislocation-twin interactions accommodate significant plasticity while maintaining extraordinary strength [5]. These phenomena suggest that the observed increase in Σ3 fraction is likely to influence strength and plasticity.

While considering the effects of boundary type, the choice of "special" boundaries should be revisited. As recognized in traditional GBE research, which boundaries are considered special depends on the property being optimized [83]. For example, corrosion and segregation resistance may be very different for the same boundary type [83]. Similar subtleties are already emerging from the study of nanocrystalline boundary networks, such as Lagrange et al. [7], who showed that a small fraction of incoherent twins can degrade the thermal stability of nanotwinned Cu. Furthermore, boundary mediated deformation mechanisms in nanocrystalline metals are only beginning to be linked to specific boundary types. Given this, it is not clear a priori which boundary types deserve the most emphasis in this study. Nonetheless, the CSL types we have focused on here are a time-tested framework, whose new implications can be further explored in future work.

**Conclusions**

The effect of deformation on nanocrystalline boundary networks has been studied using nanometer resolution orientation maps. The changes induced by monotonic and cyclic loading were quantified by analyzing the texture, grain size, grain boundary character and triple junction distributions. Deformation at room temperature did not produce microstructural evolution. Similarly, neither annealing nor monotonic loading at 100°C had any effect other than minor grain growth. Significant boundary modification was only seen under the combined influences of cyclic loading and elevated temperature. The extent of boundary evolution was dependent on the number of applied loading cycles. We conclude that the most likely mechanism is a set of collective deformation processes enabled by enhanced boundary mobility, which explains the observed temperature and cycle dependence.

In addition to providing insight into the deformation response of nanocrystalline Ni, this study suggests a pathway to improve nanocrystalline materials through grain boundary engineering. Based on the observed increase in Σ3 boundary fraction, it may be possible to use controlled plastic deformation to tailor nanocrystalline boundary networks and produce more favorable properties. Given the unique deformation processes in nanocrystalline metals, we consider it an open question what grain boundary network characteristics would be ideal. Future work may focus on linking nanocrystalline grain boundary network characteristics with different types of properties, as has been done for traditional GBE in coarse-grained materials.

**Acknowledgements**

We gratefully acknowledge support from the National Science Foundation through a CAREER Award No. DMR-1255305. This work was partly performed under the auspices of the U.S. Department of Energy by Lawrence Livermore National Laboratory under Contract DE-AC52-07NA27344. DBB and MK were supported by the U.S. Department of Energy (DOE), Office of Basic Energy Sciences, Division of Materials Science and Engineering under FWP# SCW0939. DBB also acknowledges the support of the Livermore Graduate Scholar Program (LLNL) during part of this work. TEM and TKD work was performed at the Laboratory for Electron and X-ray Instrumentation (LEXI) at UC Irvine, using instrumentation funded in part by the National Science Foundation Center for Chemistry at the Space-Time Limit (CHE-082913).

**TABLES**

**Table 1: GBCD data for mechanically cycled and annealed nanocrystalline nickel.**

| Number of Loading Cycles [time held at temperature (min)] | Temperature (°C) | Σ1 (Length %) | Σ3% (Length %) | Σ9% (Length %) | Σ27% (Length %) | Σ≤29% (Length %) | Average Grain Size (nm) |
|---|---|---|---|---|---|---|---|
| 0 | 22 | 7.4±0.9 | 23±1 | 4.3±0.7 | 0.4±0.3 | 41±2 | 22 |
| Monotonic | 22 | 7.0±0.9 | 23±1 | 4.4±0.7 | 0.5±0.3 | 42±2 | 22 |
| 125 | 22 | 6.0±1 | 23±2 | 3.2±0.7 | 1.0±0.6 | 40±2 | 22 |
| 0 [60] | 100 | 6.6±0.8 | 22±1 | 2.6±0.5 | 0.4±0.3 | 38±2 | 21 |
| 0 [79] | 100 | 5.8±0.6 | 20±1 | 3.2±0.5 | 0.7±0.3 | 37±1 | 21 |
| 0 [98] | 100 | 6.4±0.8 | 23±1 | 2.5±0.5 | 1.1±0.3 | 41±2 | 19 |
| 0 [250] | 100 | 7.1±0.7 | 22±1 | 2.5±0.4 | 0.3±0.2 | 39±1 | 27 |
| Monotonic [60] | 100 | 6.6±1 | 25±2 | 3.1±0.8 | 1.0±0.6 | 42±2 | 20 |
| 5 [79] | 100 | 5.7±0.6 | 25±1 | 3.0±0.4 | 0.4±0.2 | 41±1 | 25 |
| 10 [98] | 100 | 2.6±0.4 | 30±1 | 2.7±0.4 | 0.4±0.2 | 40±1 | 30 |
| 50 [250] | 100 | 3.7±0.4 | 34±1 | 1.9±0.3 | 0.4±0.2 | 46±1 | 51 |

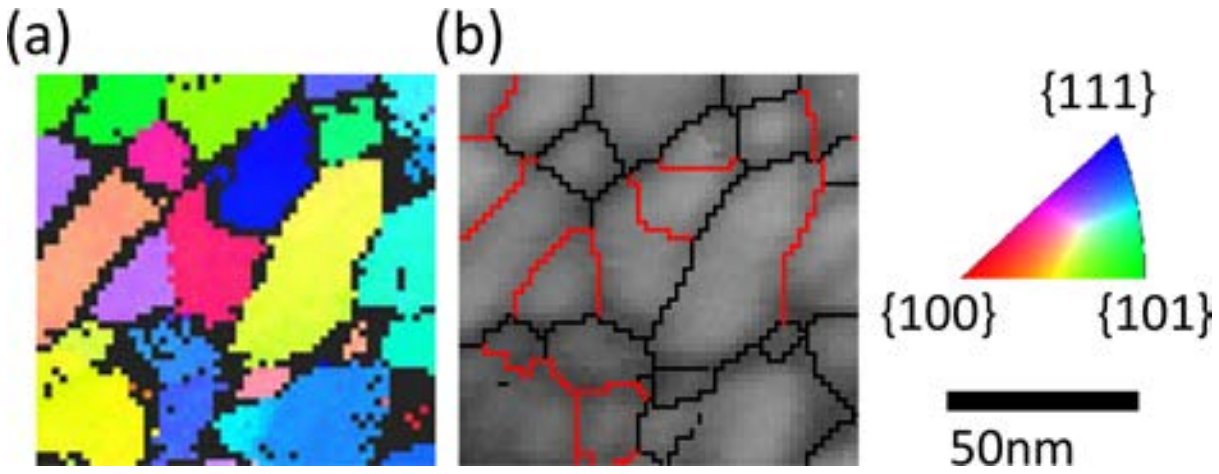

**Figure 1: (a) Raw orientation map before any post-processing. The color scheme follows the inverse pole figure legend and non-indexed pixels are colored black. (b) Reconstructed grain boundaries have been overlaid on the patter quality map. Black lines represent random boundaries, while red are Σ1-29 boundaries. Lighter shades of grey in the grain interiors indicate higher quality diffraction patterns.**

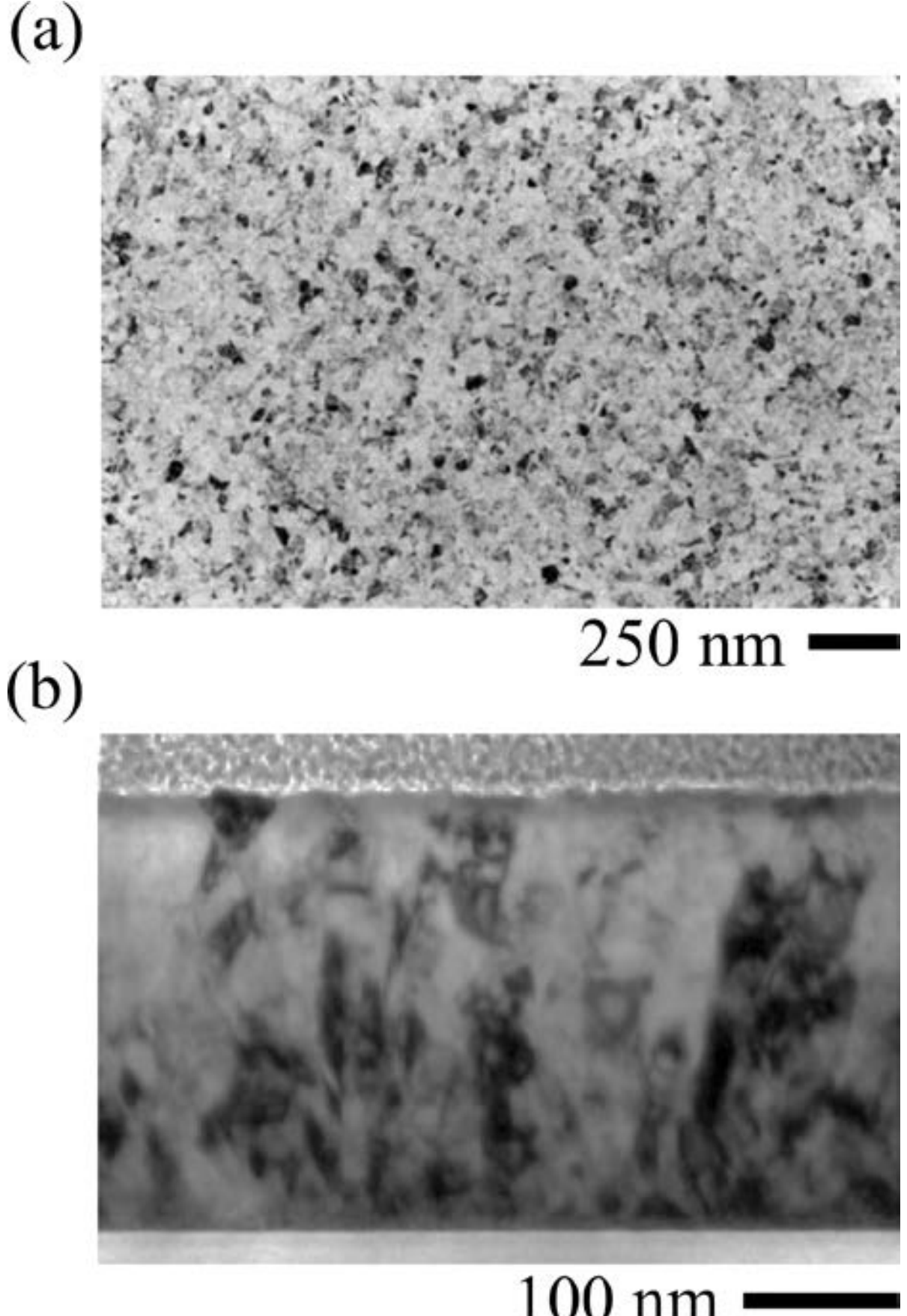

**Figure 2: Bright field TEM images of the as-deposited microstructure in (a) plan view and (b) cross sectional view.**

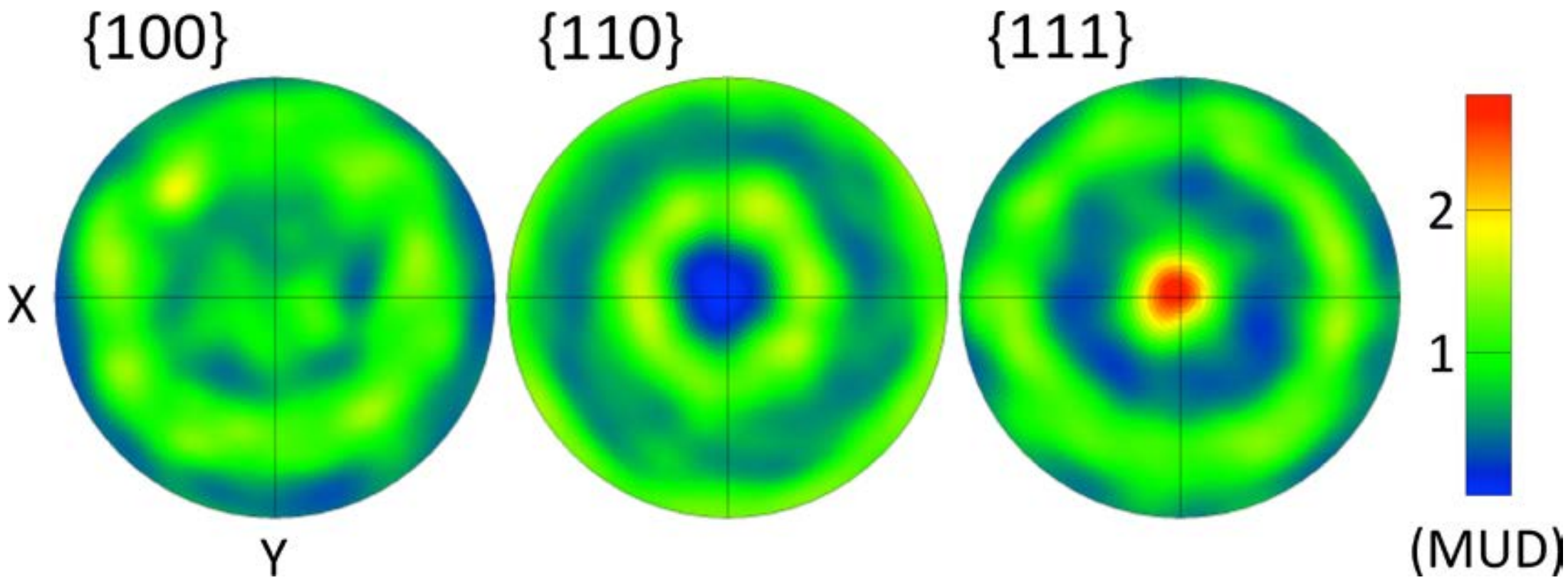

**Figure 3: Pole figure showing the texture of the as-deposited films. The color scale is in multiples of uniform distribution (MUD). Neither annealing nor mechanical cycling changed the texture.**

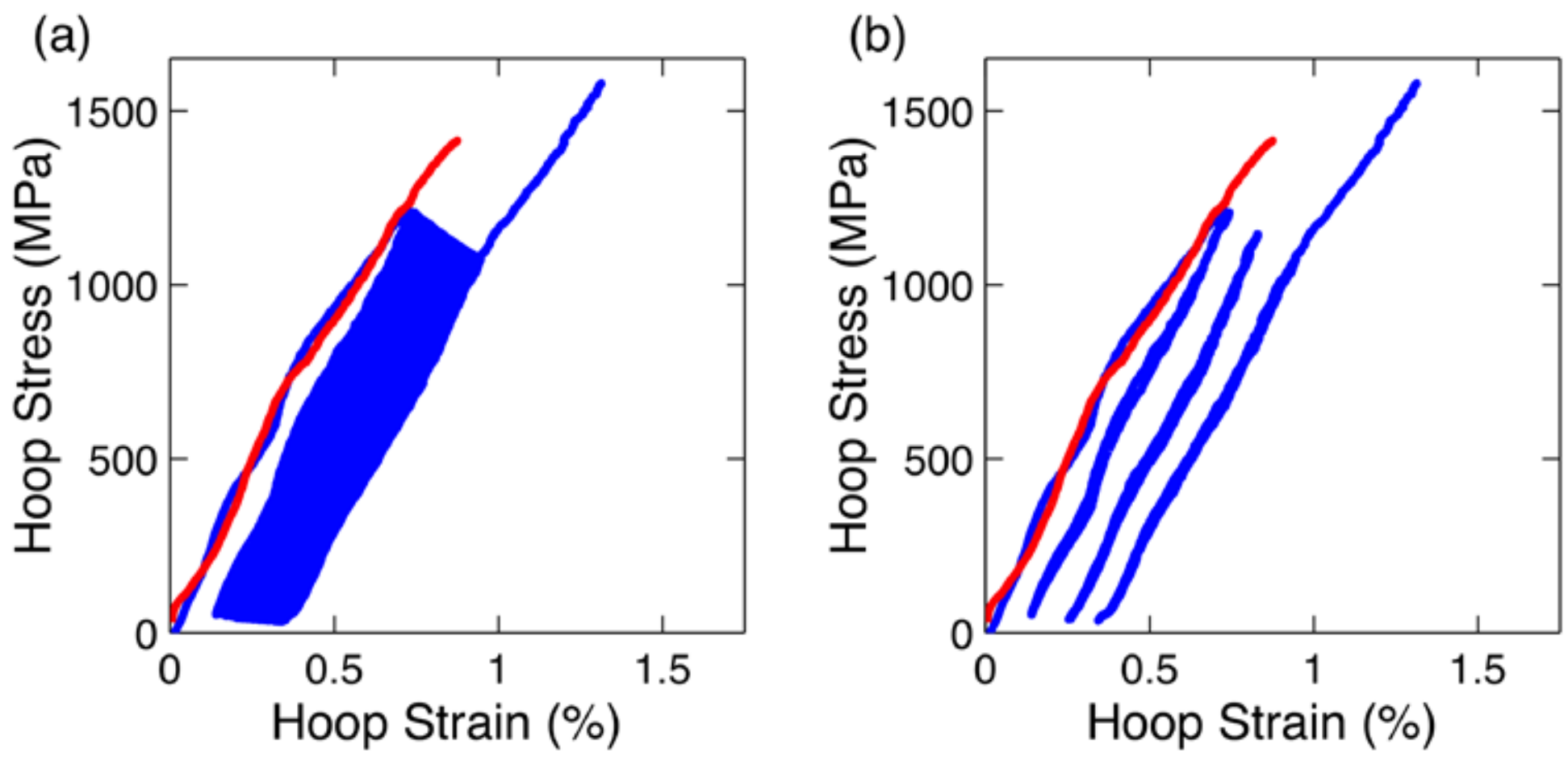

**Figure 4: Hoop stress-strain plots show the mechanical behavior of nanocrystalline nickel tested at room temperature. Part (a) shows the monotonic and full cyclic behavior, while Part (b) isolates the monotonic, 1st, 5th, 25th and 125th cycles.**

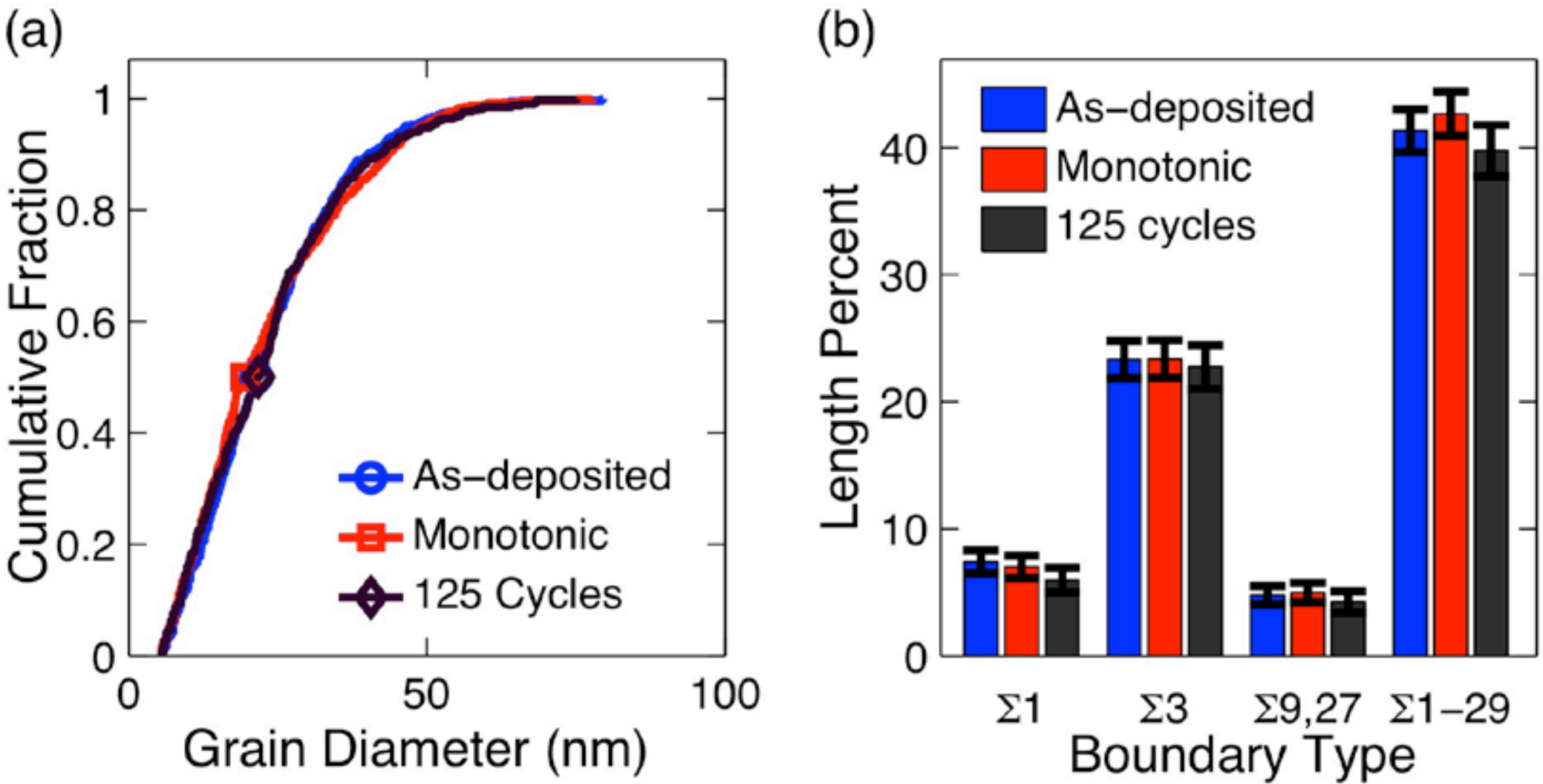

**Figure 5: Mechanical loading at room temperature had no effect on either (a) the cumulative grain size distribution or (b) the grain boundary character distribution.**

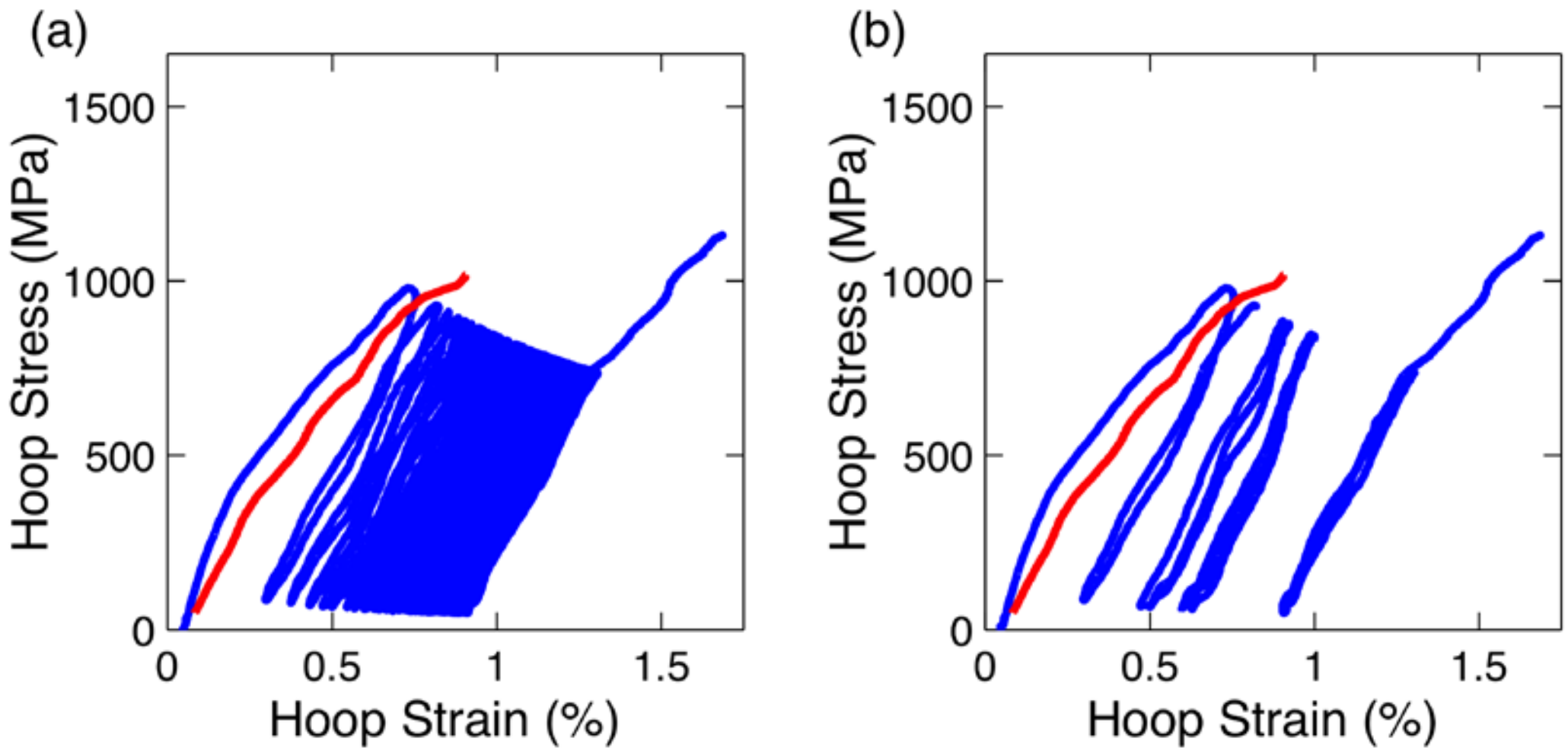

**Figure 6: Stress strain plots show the mechanical behavior of nanocrystalline nickel tested at 100 ˚C. Part (a) shows the monotonic and full cyclic behavior, while Part (b) isolates the monotonic, 1st, 5th, 10th and 50th cycles.**

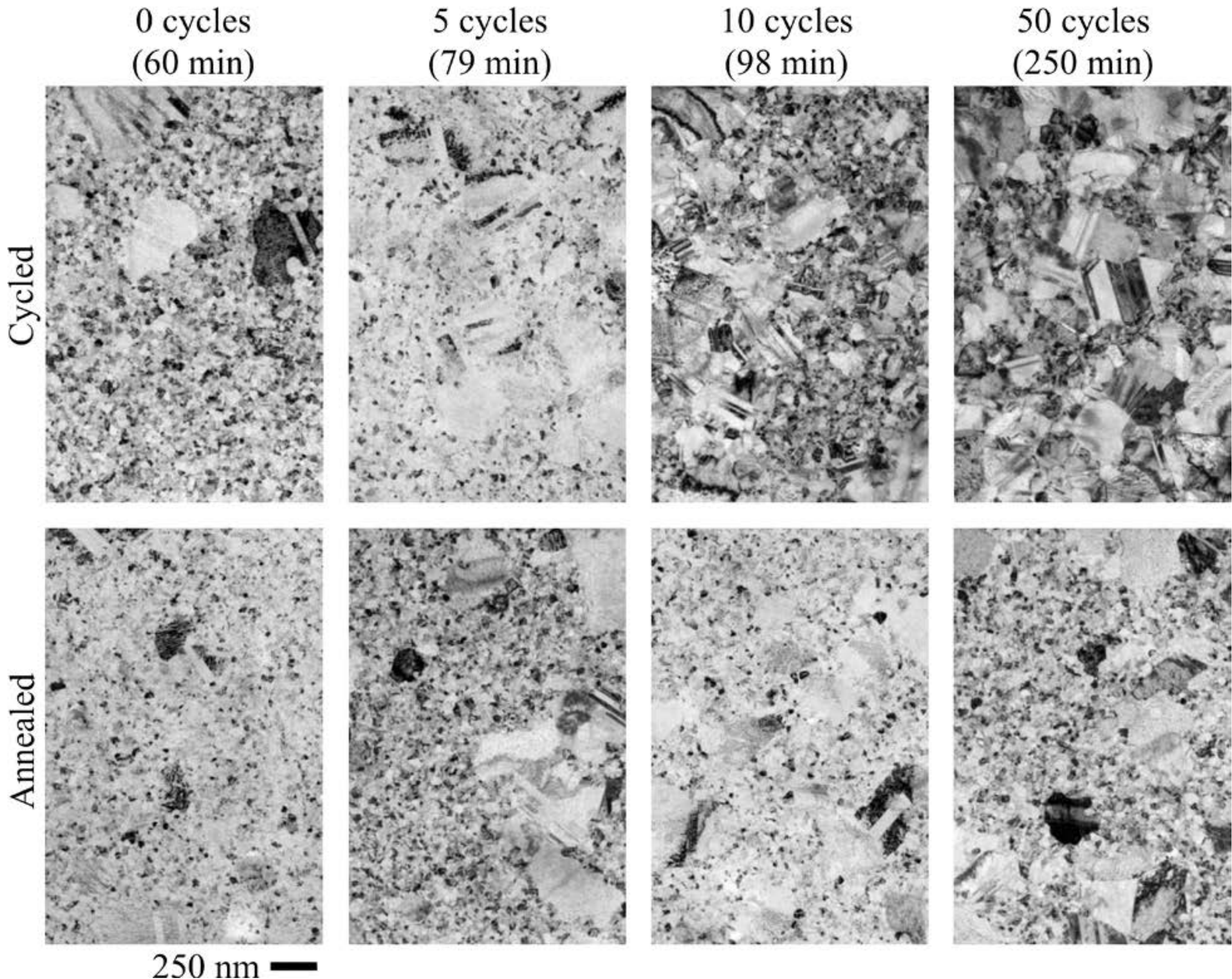

**Figure 7: The bright field TEM images in the upper row show films which were mechanically cycled at 100 °C, while those in the lower row were annealed stress-free at 100 °C for equivalent times.**

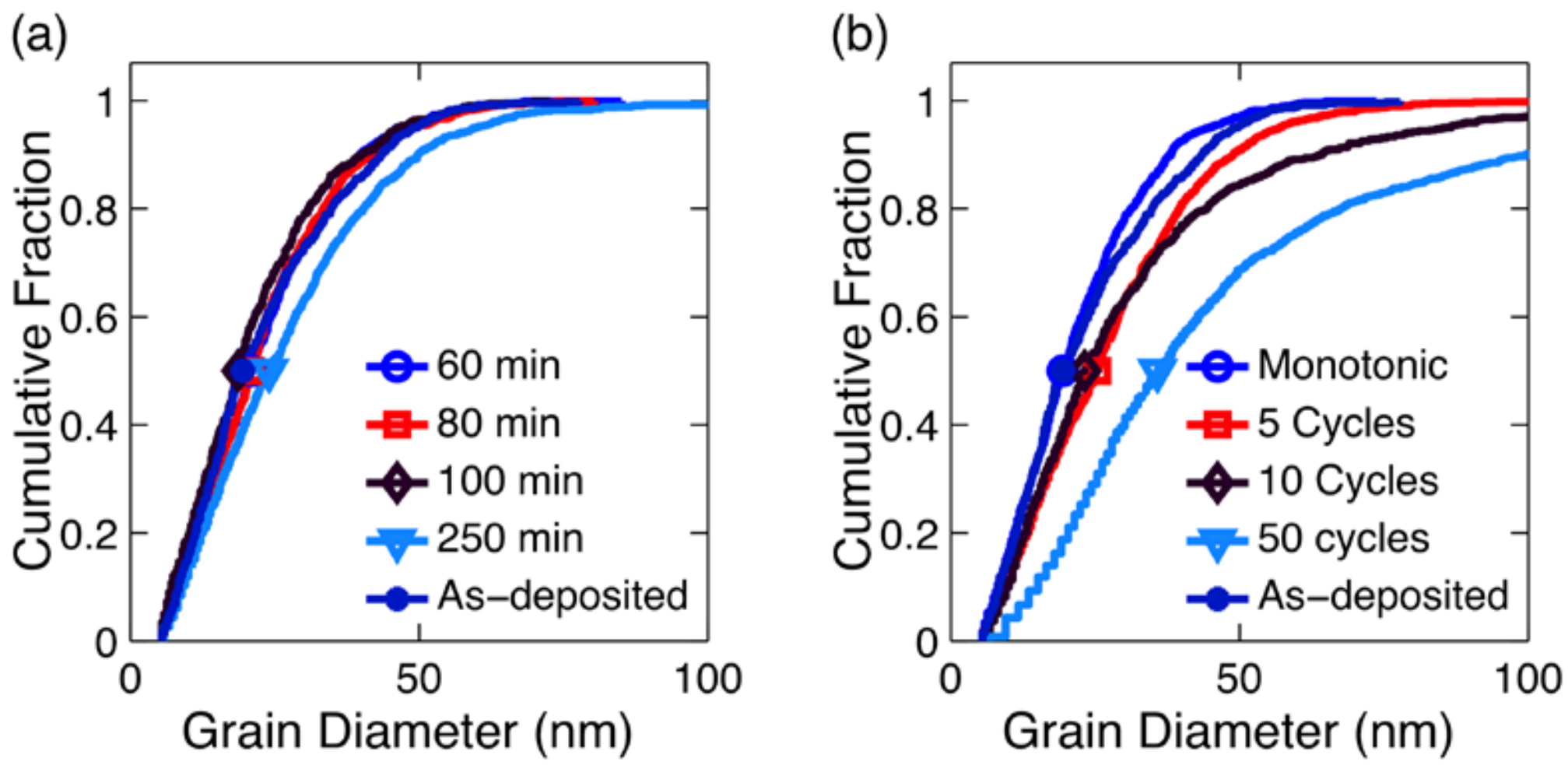

**Figure 8: (a) Stress-free annealing at 100 °C only causes subtle grain growth, while (b) mechanical cycling at 100 °C affects the cumulative grain size distribution in a much more pronounced manner.**

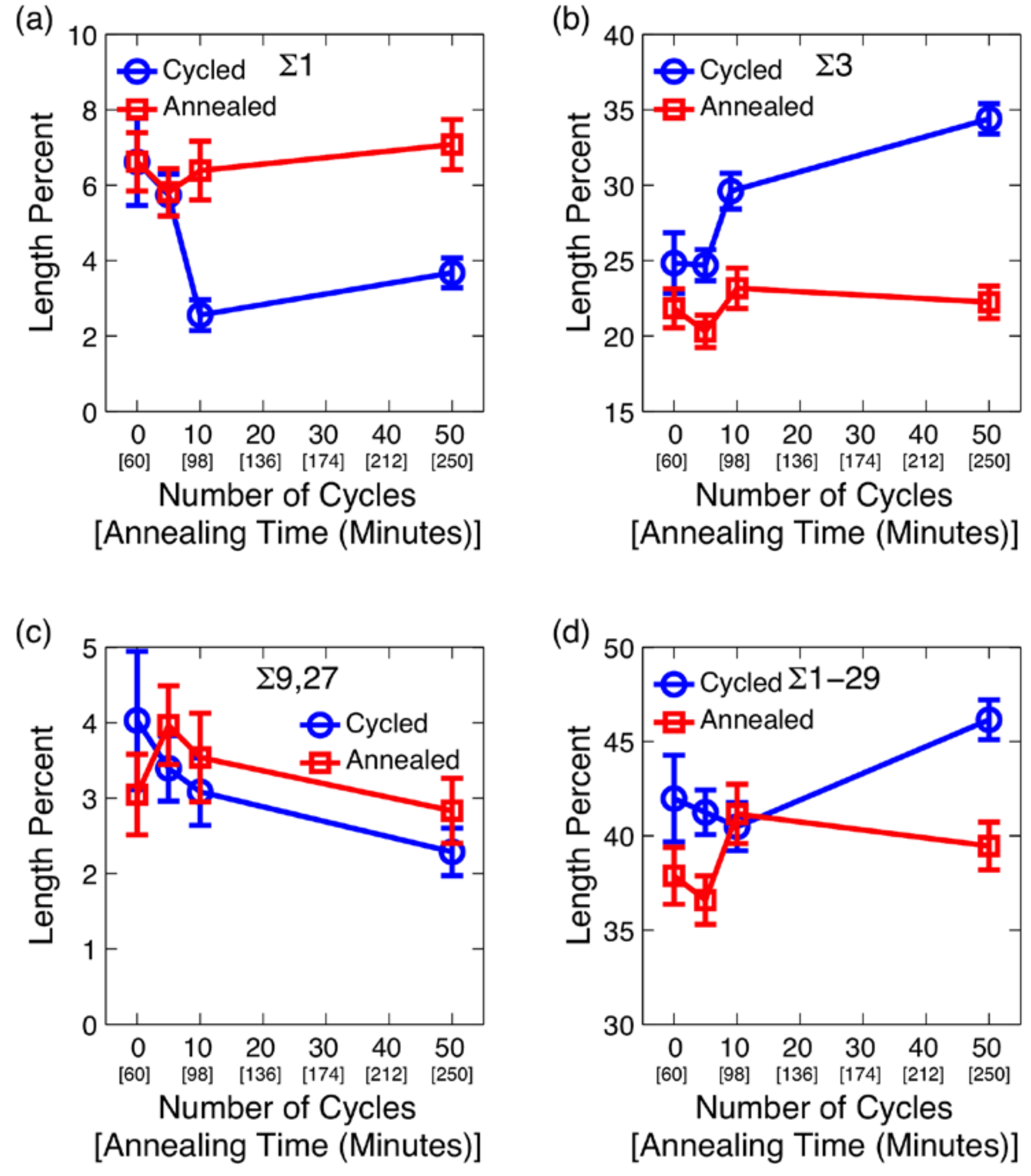

**Figure 9: The length fractions of (a) Σ1, (b) Σ3, (c) Σ9,27 and (d) Σ1-29 boundaries after stress-free annealing and mechanical cycling at 100 °C. The material was either annealed (red squares) or mechanically cycled at 100 °C (blue circles).**

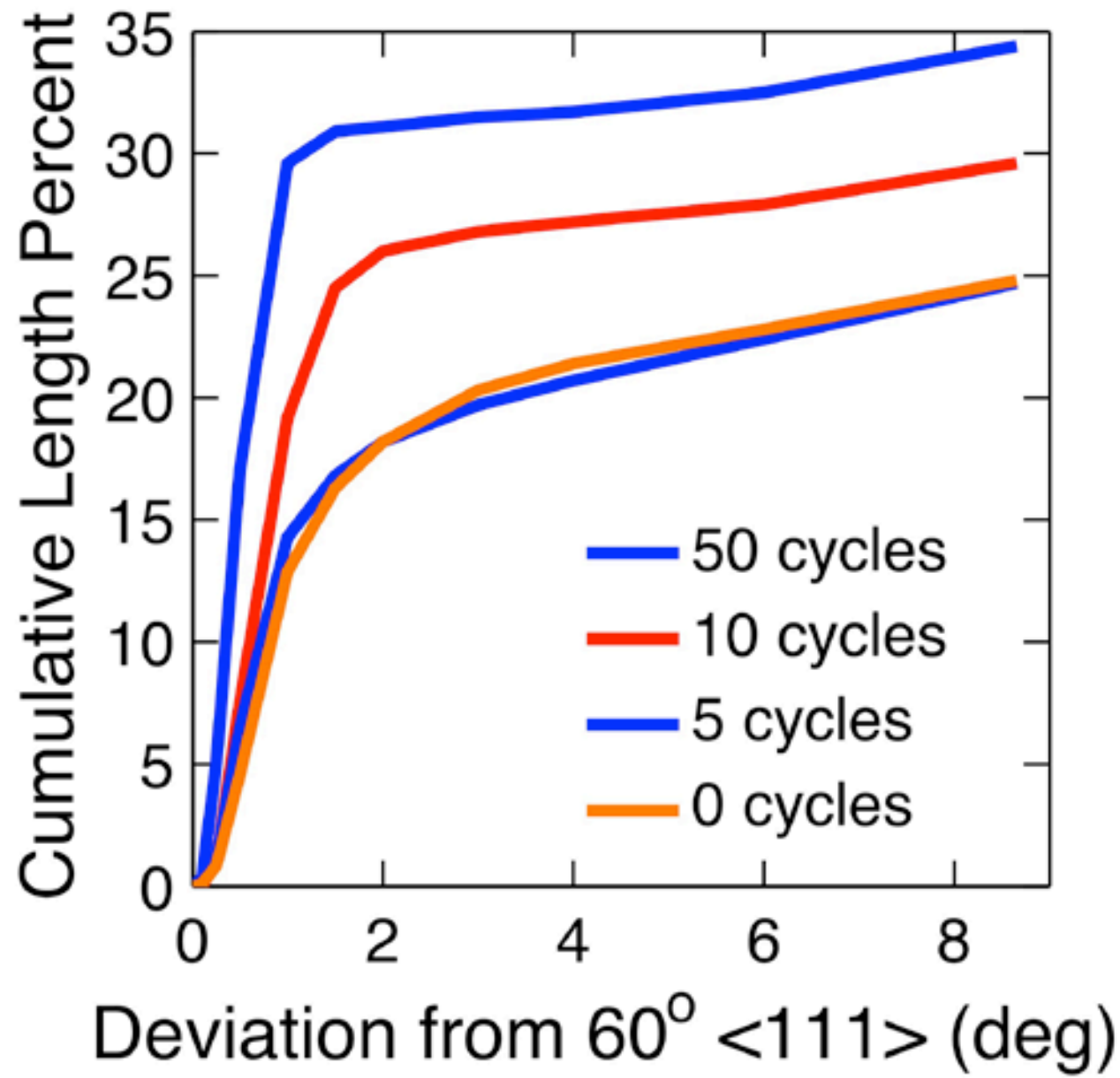

**Figure 10: The cumulative distribution of Σ3 boundaries as a function of their deviation from the ideal 60° <111> misorientation are shown for thermomechanically processed material. It shows that, on average, the Σ3 boundaries become more perfectly aligned after mechanical cycling at 100 °C.**

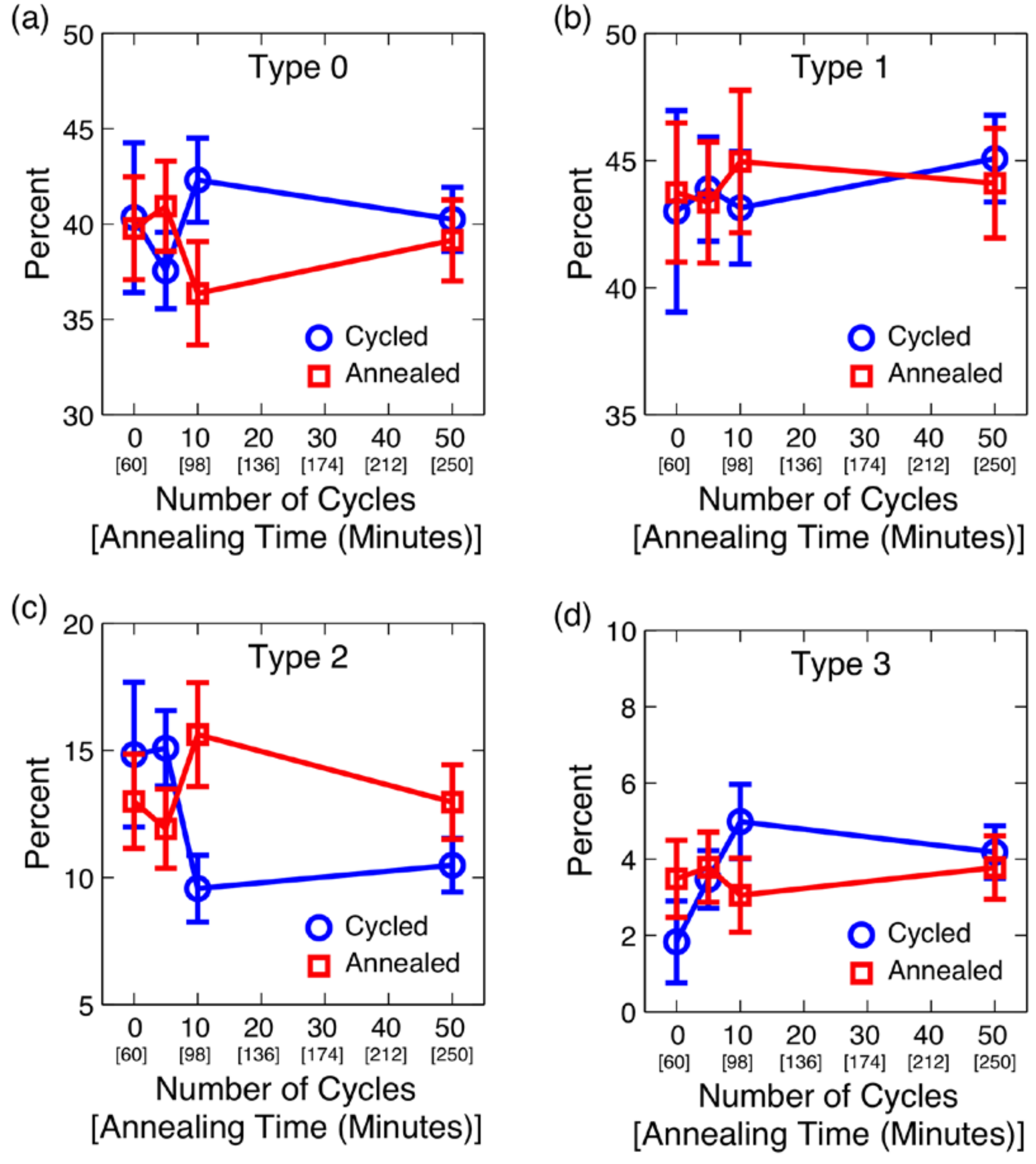

**Figure 11: The triple junction distribution is represented as the fraction of (a) type 0 (no special boundaries), (b) type 1 (1 special and 2 random boundaries), (c) type 2 (2 special and 1 random boundaries), and (d) type 3 (3 special boundaries). The material was either annealed (red squares) or mechanically cycled at 100 °C (blue circles).**

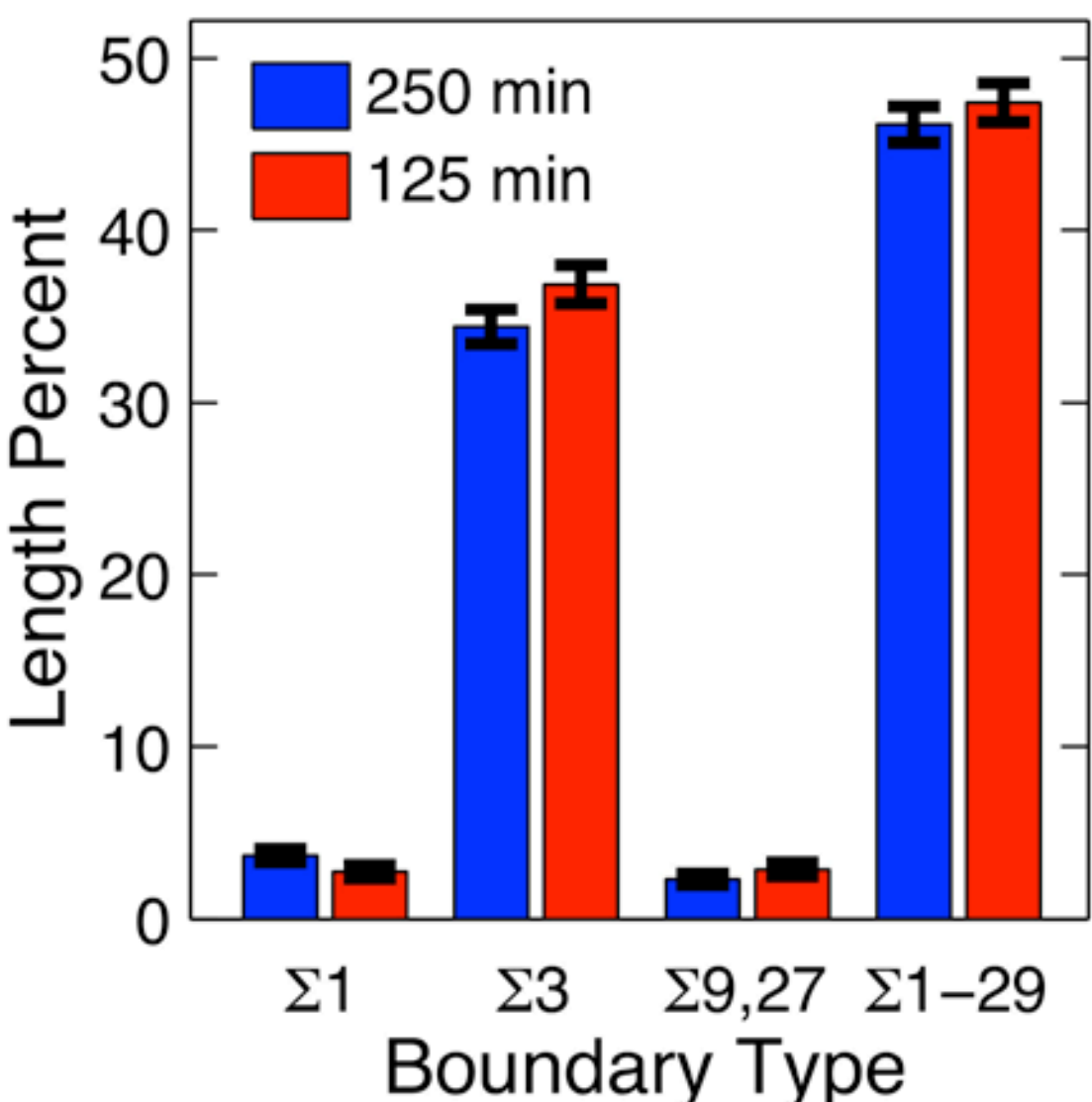

**Figure 12: For 50 mechanical cycles at 100 °C, the grain boundary character distribution is unaffected by a change from 250 to 125 minutes of total cycling duration. This indicates that creep did not contribute to the evolution of the GBCD.**